\documentclass[12pt,dvips]{article} 
\usepackage{epsfig}
\begin{document}
 
\author{
{\bf E. Cuautle, G. Herrera} 
\thanks{e-mails: gherrera@fis.cinvestav.mx,
                cuautle@fis.cinvestav.mx} \\ 
{\small Centro de Investigaci\'{o}n y de Estudios Avanzados} \\ 
{\small Apdo. Postal 14 740, M\'{e}xico 07000, DF, M\'exico}
}

\title{
Effect of residual Bose-Einstein correlations on 
the Dalitz plot of hadronic charm meson decays
}

\date{}
\maketitle

\begin{abstract}
We show that the presence of residual Bose-Einstein correlations
may affect the non-resonant contribution of hadronic charm decays 
where two identical pions appear in the final state. 
The distortion of the phase-space of the reaction would be
visible in the Dalitz plot. The decay 
$D^+ \rightarrow K^- \pi^+ \pi^+$ is discussed but results can be 
generalized to any decay with identical bosons.
\end{abstract}

\newpage


\section{Introduction}


Hadronic decays of D mesons are an important source of information.
The study of pseudoscalar-pseudoscalar and pseudoscalar-vector decays
of D mesons, for example, shed light on heavy quark decay and 
hadronization.\\
The lifetime difference between the $D^+$ and the $D^0$ mesons
as well as the role of final state interactions may be 
understood by carefully studying their hadronic decays.\\

The fractions of resonant (pseudoscalar-pseudoscalar or pseudoscalar-vector)
and non resonant three-body contributions to each decay mode are 
determined by fitting the Dalitz plot distributions to a coherent sum of 
amplitudes.\\
The three-body amplitude is normally  assumed to be constant 
over the Dalitz plot but, decays like $D^+ \rightarrow K^- \pi^+ \pi^+$ 
contain identical charged pions in the final state and their interference 
produce a non-constant behaviour of the phase-space \cite{markiii}.\\
The first evidence for a non-uniform population on the Dalitz plot for
$D^+ \rightarrow K^- \pi^+ \pi^+$ was shown in ref. \cite{schindler}.\\
In \cite{markiii} an attempt is made to account for the interference
by adding an ad hoc term to the likelihood function. The best fit is
obtained with a Bose symmetric term of the form $B_{3 body} = \mid 
m^2_{K\pi1} - m^2_{K\pi2} \mid $. However, they found that this form does 
not seem to adequately describe the physics.\\ 
In a recent analysis \cite{e687} the amplitudes of the fit were again 
unsuccessfully Bose-symmetrized in an attempt to describe the 
non-resonant contribution.\\

Recently Bediaga {\it et al.} \cite{bediaga} studied the problem 
and found that the non-resonant contribution should actually not 
be constant over the phase-space. They used a method based on
factorization and an effective hamiltonian for the partonic interaction 
to estimate the distortion of the Dalitz plot.\\

Here we will try to address the problem from the perspective of Bose-Einstein
interference among the identical bosons in the final state and the
residual correlation produced. This may help to understand
the Dalitz plot of $D^+ \rightarrow K^- \pi^+ \pi^+$ in particular, 
and the effect in any decay with several bosons in general.

 
\section{Bose-Einstein correlations} 


The study of Bose-Einstein correlations (BEC) in connection
with other phenomena \cite{residual} in high energy physics reactions
became important recently when it was realized that such correlations may 
affect measurements of the standard model parameters 
\cite{bialas,sjostrand}.\\

Charm mesons travel long enough to neglect the effect of interference
among their decay products and the pions produced in the reaction.
Effects of the kind as described in \cite{residual} will not be
considered here. We rather will regard the decay itself as a particle 
production process in which the interference may arise.\\

Fig. 1(a) shows the decay with a bubble representing the region in space-time
where particles are produced. Decays where particles are produced 
via fragmentation are not completely coherent. The incoherence present in 
the decay gives rise to the Bose -Einstein interference.\\

The Bose-Einstein correlations are commonly described in terms of a two
particle correlation function:  
\begin{equation}
 R_{BE}(p_1,p_2) = \frac{P(p_1,p_2)}{P(p_1)P(p_2)}
\end{equation}
\noindent
where $P(p_1,p_2)$ is the joint probability amplitude for the emission of
two bosons with momenta $p_1$ and $p_2$, and $P(p_1)$ and $P(p_2)$ 
are the single production probabilities.\\

The Bose-Einstein correlation among identical mesons has been used to 
probe the space-time structure of the intermediate state right before
hadrons appear \cite{gold,us} in high energy and nuclear collisions.\\

Real data are analysed in terms of the ratio of distributions of like
charged pairs (identical bosons) to that of an uncorrelated sample
\begin{equation}
 R_{BE}(p_1,p_2) = \frac{C(p_1,p_2)}{C_0(p_1,p_2)}.
\end{equation}
The uncorrelated sample $C_0(p_1,p_2)$ with all the features of $C(p_1,p_2)$
except for the Bose-Einstein effect may be obtained from the distribution
of unlike charged pions or by using an event mixing technique.\\

One parametrizes the effect assuming a set of point-like sources emmiting
bosons. These point like sources are distributed according to a 
density $\rho (r)$. The correlation function can then be written as, 
\begin{equation}
 R_{BE}(\vec{p_1},\vec{p_2}) = \int \rho (\vec{r_1}) \rho (\vec{r_2}) 
\mid \psi_{BE}(\vec{p_1},\vec{p_2}) \mid ^2 d^3 r_1 d^3 r_2,
\end{equation}
\noindent
where $\vec{p_1},\vec{p_2}$ are the momenta of the two bosons, 
$\psi_{BE}$ represents the Bose-Einstein symmetrized wave function
of the bosons system and $\int _V \rho (\vec{r}) d^3 r = 1 $.
Taking plane waves to describe the bosons one obtains
the correlation function for an incoherent source:
\begin{equation}
 R_{BE}(\vec{p_1},\vec{p_2}) = 1 + \mid {\cal F} ( \rho(\vec{r}) ) \mid ^2,
\end{equation}
\noindent
where $\cal F (\rho)$ represents the Fourier transform of the density
function $\rho(\vec{r})$.\\

In order to describe the quantum interference during the fragmentation
in high energy reactions, phenomenological parametrizations of the effect
have been proposed. For a recent review see \cite{boal}. 

One of the most commonly used parametrizations is given by:
\begin{equation}
 R_{BE}(\vec{p_1},\vec{p_2}) = 1 + \lambda e^{-Q^2 \beta},
\end{equation}
where $Q^2$ is the Lorentz invariant, $Q^2 = -(p_1-p_2)$, which can be
written also as, $Q^2 = m^2 - 4 m_{\pi}^2 $, where $m$ is the invariant
mass of the two pions and  $m_{\pi}=0.139 GeV$ the mass of the pion.
The parameter $\lambda$ lies between 0 and 1 and reflects the degree
of coherence in the pion production. The radius of the pions
source will be given by $r = \hbar c \sqrt{\beta} [fm]$.

The presence of Bose-Einstein correlations will modify not only the
invariant mass spectrum of like charged but also that of unlike
charged pions. This reflection of BEC known as residual correlation 
has been studied since long time ago to make sure that the reference 
sample of unlike charged pions used to subtract the effect from the like 
charged pions spectra is clean.\\
In particle production processes with high multiplicity, residual
correlations
easily dissolve. Altough some studies \cite{residual} claim that 
this is not the case and that using unlike charged pions as a reference
sample is questionable.\\
In a particle production process where only three particles are yield
({\it e.g.} $D^+ \rightarrow K^- \pi^+ \pi^+$)  residual correlations are 
expected to be larger.


\section{Simulation of Bose-Einstein correlations}


In this letter we want to estimate the effects of BEC on the phase space 
of a three body decay. We take the particular case of 
$D^+ \rightarrow K^- \pi^+ \pi^+$ and look at it, as a particle
production process with only two identical bosons in the final state. The 
reduced number of pions makes easy the incorporation of BEC in a Monte 
Carlo simulation. We will take the approach used in \cite{residual} where
the BEC are simply simulated by weighting each event. The MC 
generator consists of a three body decay simulation with the appropiate 
masses of the decay products. In \cite{residual} each event
is weighted according to:
\begin{equation}
W=\prod_{i,j} (1+ \lambda e^{-\beta Q_{ij}^2}),
\end{equation}
where the product would be taken over all pairs $(i,j)$ of like charged 
pions.\\
In our case the product is reduced to one pair of like charged pions
\begin{equation}
W= 1+ \lambda e^{-\beta Q^2}.
\end{equation}
We did simulations with different values for $\lambda$ and $\beta$.
Fig. 2 shows the squared invariant mass distributions both for $K^-\pi^+$
and $\pi^+\pi^+$ before and after simulation of BEC. 
The $Q^2(\pi^+ \pi^+)$ distribution can be normalized using its
distribution in the abscence of BEC. Fig. 3 shows the correlation function
$R_{BE}(\pi^+ \pi^+)$ obtained according to eq. (2) and the curve of eq. (5) 
with $\lambda=1$ and $\beta=4 GeV^{-2}$. 
One can see that one obtains from the simulation exactly what one puts in.\\


\section{Effects on the Dalitz plot}


The effect of BEC on the Dalitz plot in some cases
would be a straightforward manifestation of the interference among
identical pions. An example of this is the Dalitz plot of $D^+ 
\rightarrow K^- \pi^+ \pi^+$ where the invariant mass squared of the two
pions is plotted agains the invariant mass squared of the kaon and one of
the pions,
{\it i.e.} $m^2(\pi \pi)$ {\it vs.} $m^2(K \pi)$. In this case the BEC
will be visible directly in $m^2(\pi \pi)$. The invariant mass spectra of
$m^2(K \pi)$, however, will show residual effects.\\
The Dalitz plot of $m^2(K \pi_1)$ {\it vs.} $m^2(K \pi_2)$ will show only 
residual effects.\\

Fig. 4 shows the Dalitz plots for the  $D^+ \rightarrow K^- \pi^+ \pi^+$
decay once the BEC has been incorporated. The values $\lambda=1$ and 
$\beta=4 GeV^{-2}$ were used to produce the distributions shown.
Increasing $\beta$ from $4 GeV^{-2}$ to $100 GeV^{-2}$ would make the 
long bands in Fig. 4 narrower. As mentioned above, the physical
meaning of $\beta=4,9,16,25,100 GeV^{-2}$ is given in terms of 
the pions source radius $r \approx  0.4,0.6,0.8,1.0,2.0 fm$ respectively.
We show the results using $r = 0.4 fm$ just because the effect is more 
visible.
Of course the real value would be extracted from fits to experimental 
data.\\
A better fit of the non-resonant part of the decay would be possible
once  $r$ has been extracted from data.\\

As in the case of the parameter $\beta$, we used $\lambda=1$ 
because this produces the biggest possible effect. The value 
$\lambda=1$ indicates a totally chaotic production in the decay. 
Of course a particle decay is not necessarily a completely chaotic process. 
In fact, if fragmentation would not take place during the decay, one may
expect a completely coherent process in which $\lambda=0$ and
Bose-Einstein interference  would not be present. Fragmentation in the
decay introduces some degree of incoherence giving to $\lambda$ a value
between 0 and 1.
The exact value would be obtained fitting the correlation 
function as in the case of the source radius. On the other hand,
there may be production mechanisms other than fragmentation that introduce
some degree of incoherence. The study of BEC in particular decays may
help to disentangle the production process.\\

The hadronic decay $D^+ \rightarrow K^- \pi^+ \pi^+$ seems to be dominated by 
the non resonant contribution \cite{pdg} and is therefore an interesting 
laboratory to study the role of BEC in hadronic decays in general. 
A resonant decay like $D^+ \rightarrow \bar{K^{*0}} \pi^+$ corresponds
to a situation illustrated in Fig. 1(b). The interference between
the pions in this case would, according to Bowler \cite{bowler}, 
reduce and narrow the strength of the correlation.\\
In hadronic decays where the resonant contribution is significant such
effects must be taken into account.\\  

An interesting aspect of the fact that BEC correlations are observable 
in hadronic decays of charm and eventually beauty mesons is that
there must be some relationship between the parametrization of $\rho(r)$
for the pions source and the D meson form factors.
Such relationship can be examined and the study of BEC may be a tool
to learn more about the decay and fragmentation mechanism.


\section*{Acknowledgments} 


We wish to thank Ignacio Bediaga, Ramon M\'endez Galain and Carla G\"obel
from Centro Brasileiro de Pesquisas F\'{\i}sicas (CBPF), Brazil,
for the very inspiring  discussions.


\newpage

\section*{Figure Captions}
 
\begin{itemize}
\item 
[Fig. 1:] Picture of (a) the non-resonant part of the decay
in which the partonic processes are overlooked. The bubble
described by $\rho(r)$, parametrizes the space time of the production
mechanism. In (b) a resonant decay is illustrated.

\item 
[Fig. 2:] Squared invariant mass distributions for $K^-\pi^+$
and $\pi^+\pi^+$  with (histogram) and without (crosses) BEC.

\item 
[Fig. 3:] Correlation function $R_{BE}$ obtained normalizing the 
$Q^2(\pi^+ \pi^+)$ distribution in the presence of BEC with the 
$Q^2(\pi^+ \pi^+)$ 
distribution without BEC. The curve corresponds to eq. (5) 
with $\lambda=1$ and $\beta=4 GeV^{-2}$.

\item 
[Fig. 4:] Dalitz plots for the  $D^+ \rightarrow K^- \pi^+ \pi^+$
decay once the BEC has been incorporated.\\

\end{itemize}

\newpage

\begin{figure}[b] 
\epsfig{figure=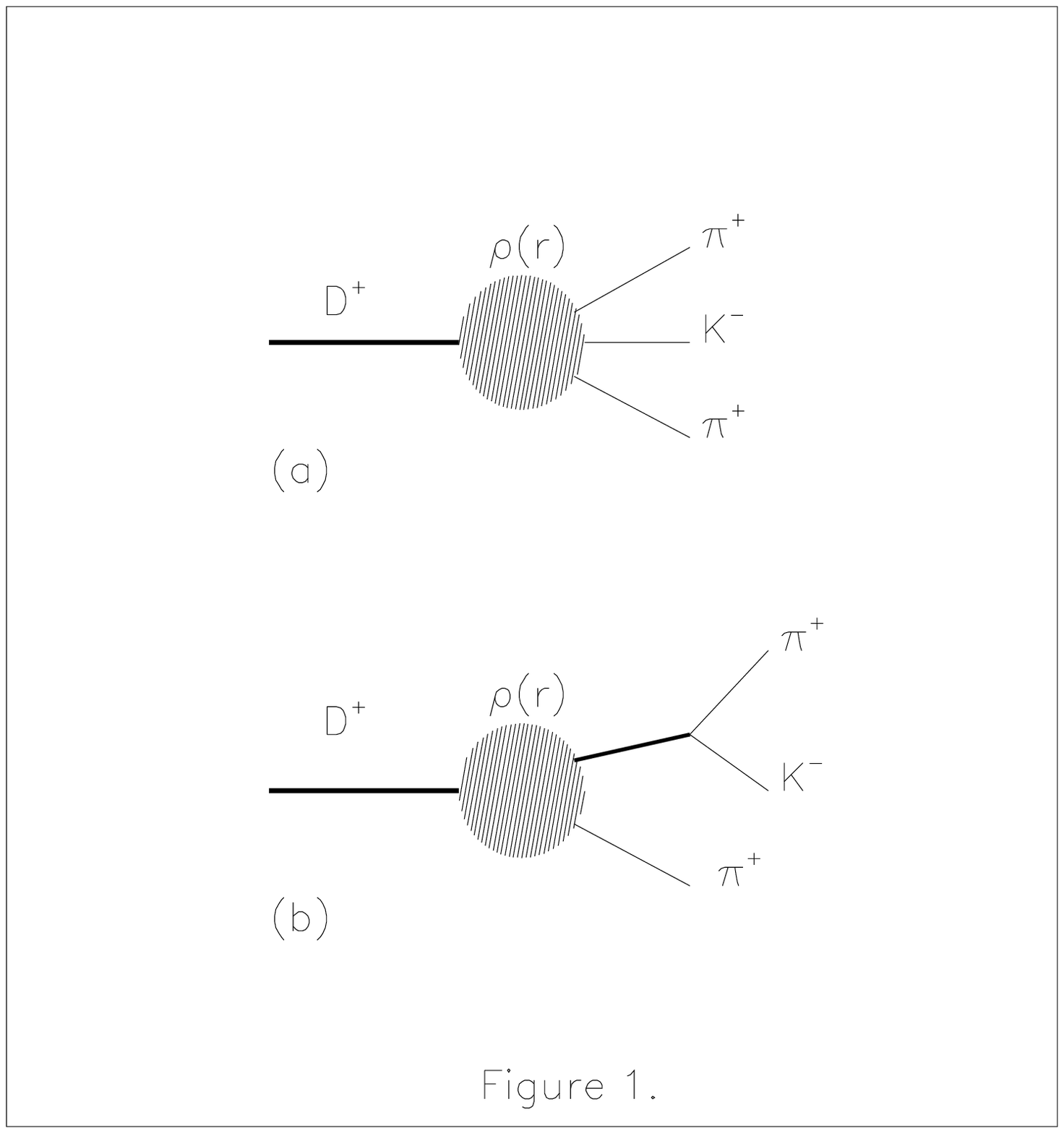,height=6.0in}
\end{figure} 

\begin{figure}[b] 
\epsfig{figure=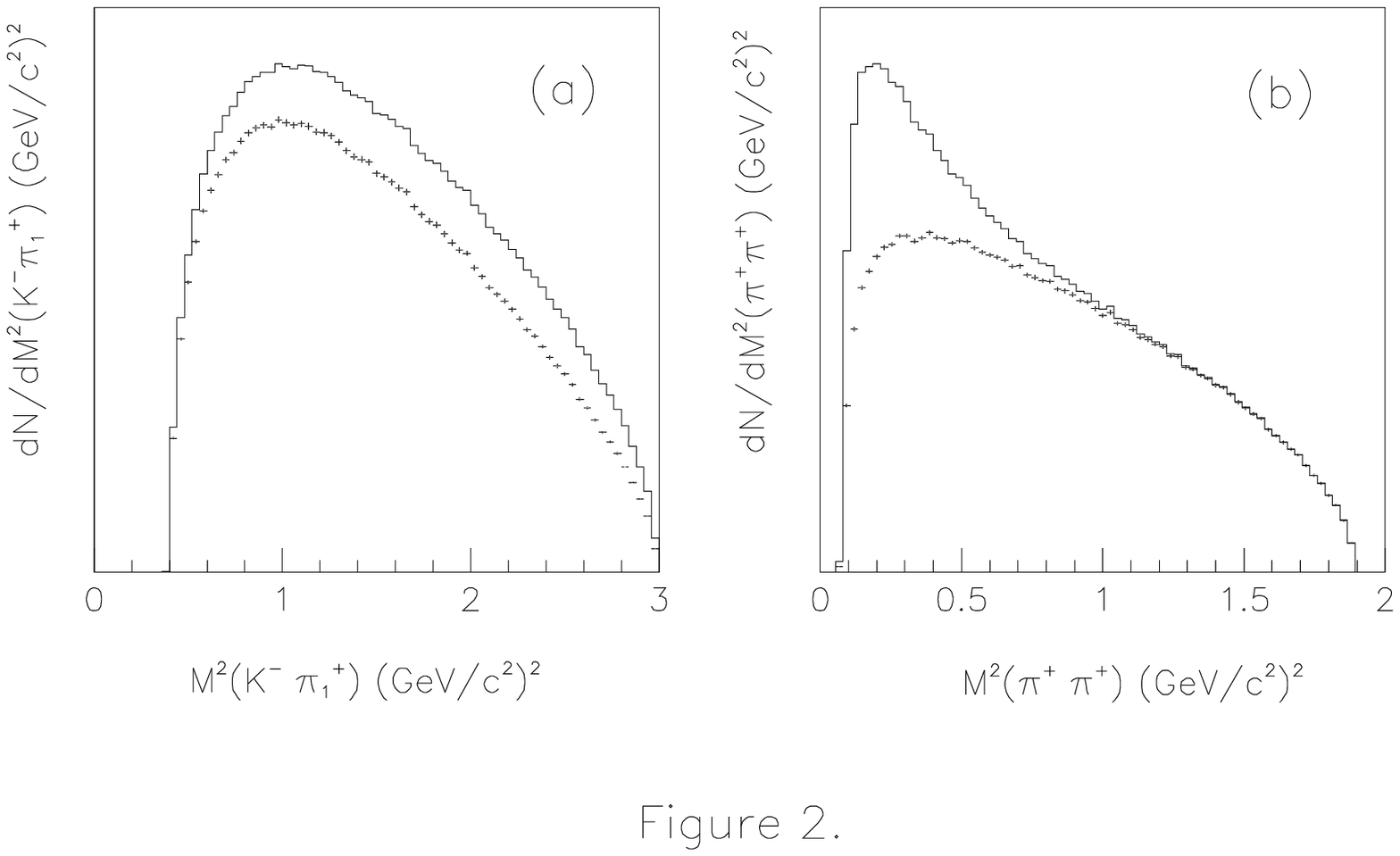,height=6.0in}
\end{figure} 

\begin{figure}[b] 
\epsfig{figure=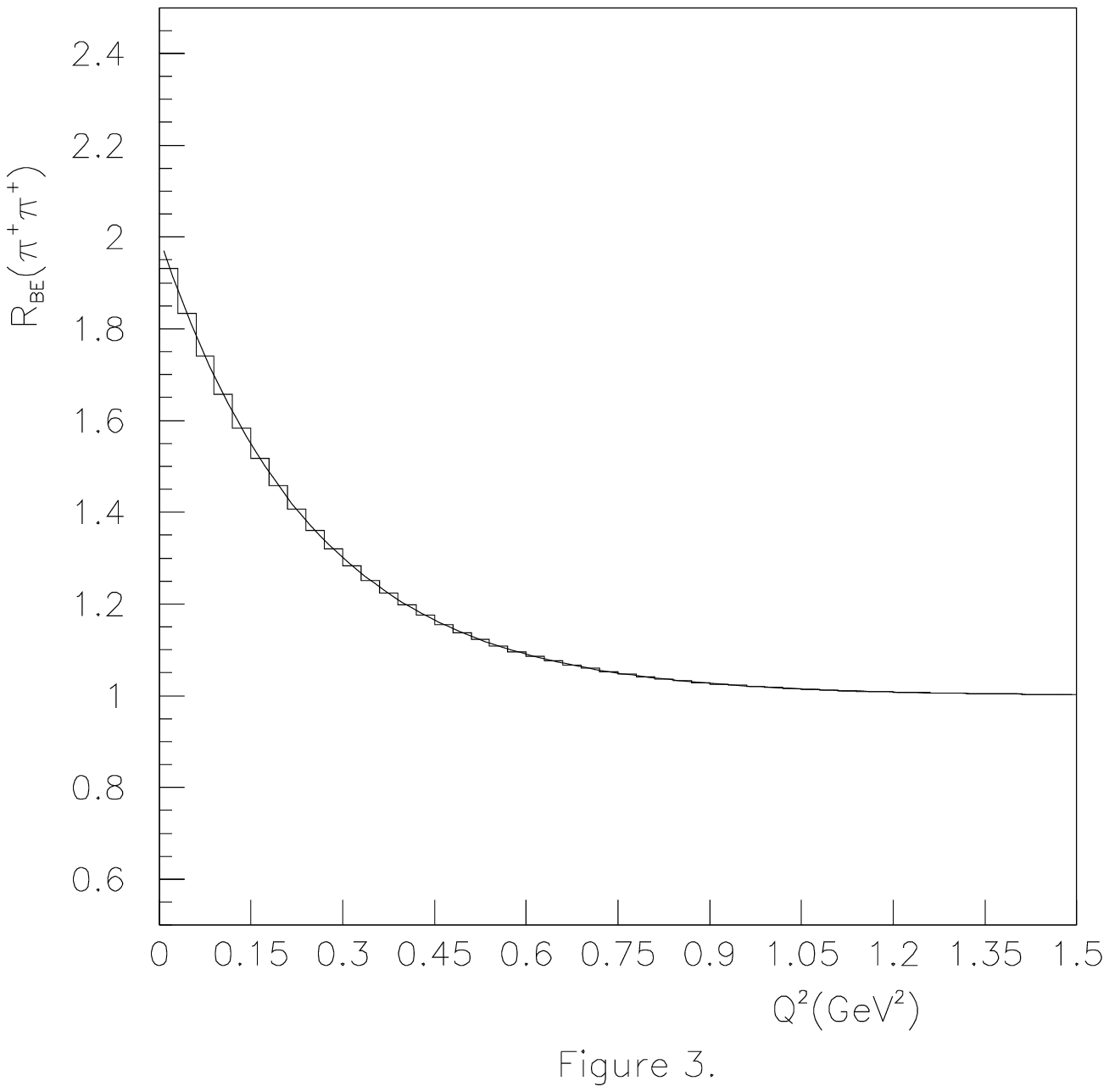,height=6.0in}
\end{figure} 

\begin{figure}[b] 
\epsfig{figure=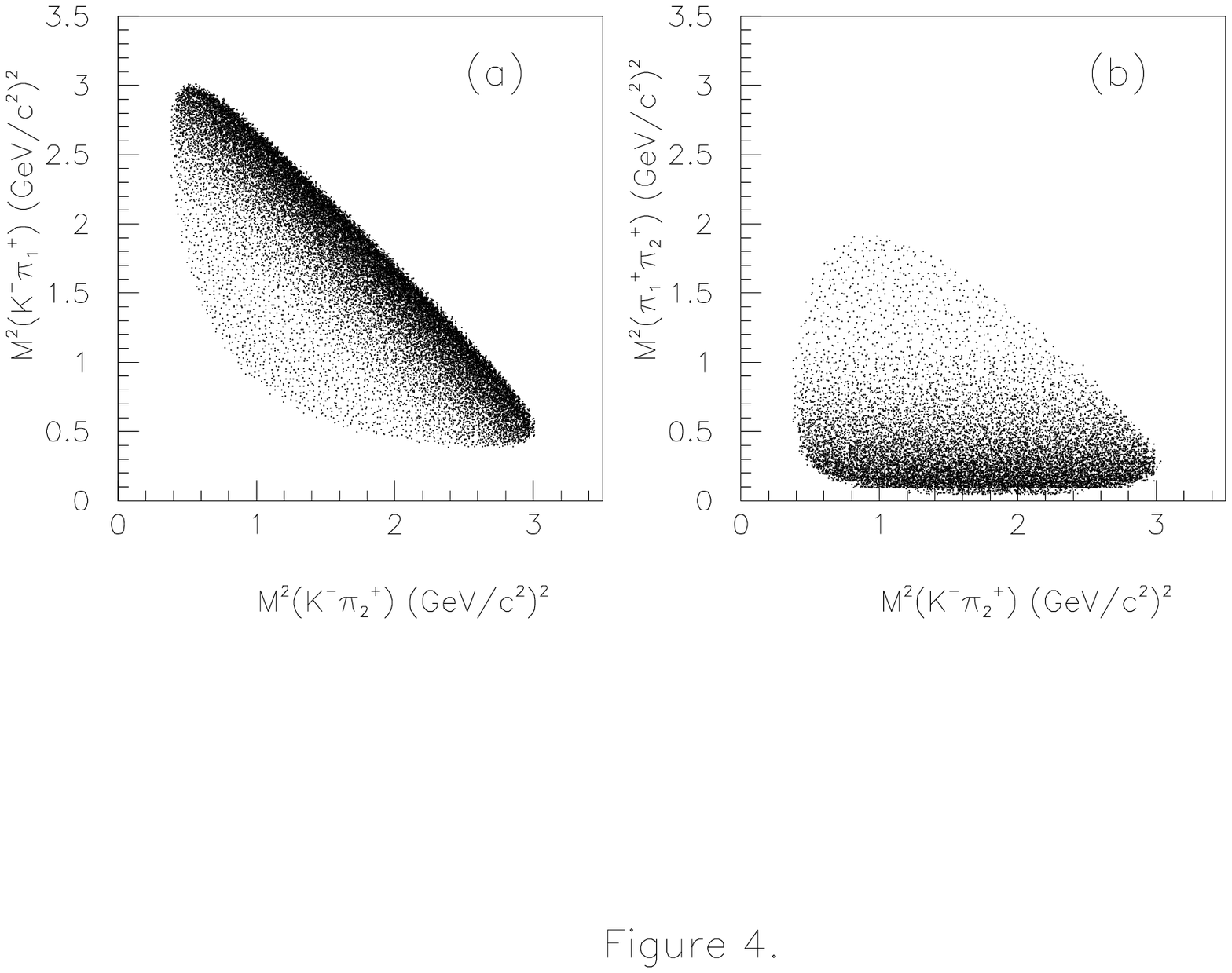,height=6.0in}
\end{figure}

\end{document}